\documentclass{nature}

\usepackage{color}
\usepackage{siunitx}
\usepackage{lineno}
\usepackage{graphicx}
\usepackage{float}
\usepackage{caption}
\usepackage[hidelinks]{hyperref}
\usepackage{cleveref}

\title{Uniformly hot nightside temperatures on short-period gas giants}

\begin{document}
\author{Dylan Keating$^{1,2,3}$, Nicolas B. Cowan$^{1,2,3,4}$, Lisa Dang$^{1,2,3}$}

\maketitle

\begin{affiliations}
\item{Department of Physics, McGill University, 3600 rue University, Montr\'eal, QC, H3A 2T8, CAN}
\item{McGill Space Institute (MSI), McGill University, 3550 rue University, Montr\'eal, QC H3A 2A7, CAN}
\item{Institut de Recherche sur les Exoplan\`etes (iREx), Universit\'e de Montr\'eal, C.P. 6128 Succ. Centre-ville, Montr\'eal, QC H3C 3J7, CAN}
\item{Department of Earth \& Planetary Sciences, McGill University, 3450 rue University, Montr\'eal, QC, H3A 0E8, CAN}
\end{affiliations}

\textbf{
Short-period gas giants (hot Jupiters) on circular orbits are expected to be tidally locked into synchronous rotation, with permanent daysides that face their host stars, and permanent nightsides that face the darkness of space \cite{Showman2002}. Thermal flux from the nightside of several hot Jupiters has been measured, meaning energy is transported from day to night in some fashion. However, it is not clear exactly what the physical information from these detections reveals about the atmospheric dynamics of hot Jupiters. Here we show that the nightside effective temperatures of a sample of 12 hot Jupiters are clustered around 1100 K, with a slight upward trend as a function of stellar irradiation. The clustering is not predicted by cloud-free atmospheric circulation models \cite{Komacek2017,Komacek2018,ZhangShowman2017}. This result can be explained if most hot Jupiters have nightside clouds that are optically thick to outgoing longwave radiation and hence radiate at the cloud-top temperature, and progressively disperse for planets receiving greater incident flux. Phase curve observations at a greater range of wavelengths are crucial to determining the extent of cloud coverage, as well as the cloud composition on hot Jupiter nightsides \cite{Morley2017,Tinetti2018}.} 

We collected published full orbit, infrared phase curves for twelve hot Jupiters: CoRoT-2b \cite{Dang2018}, HAT-P-7-b \cite{Wong2016}, HD 149026b \cite{Zhang2018}, HD 189733b \cite{Knutson2012}, HD 209458b \cite{Zellem2014}, WASP-12b \cite{Cowan2012}, WASP-14b \cite{Wong2015}, WASP-18b \cite{Maxted2013}, WASP-19b \cite{Wong2016}, WASP-33b \cite{Zhang2018}, WASP-43b \cite{Stevenson2014,Stevenson2017,Mendonca2018,Spiderman}, and WASP-103b \cite{Kreidberg2018}. We also included the brown dwarf KELT-1b \cite{Beatty2018}. We calculated the nightside brightness temperatures from the phase curve parameters, and used Gaussian Process regression to estimate each planet's bolometric flux, and subsequently its disk-integrated nightside effective temperature. Several of the published phase curve fits imply negative nightside disk-integrated flux, which is unphysical, because it implies that the planets have negative brightness at some longitudes on their surface. We explain how we handled these cases in the Methods section. Future phase curve observations should be fit with the constraint that flux is non-negative everywhere on the planet. We also inferred nightside temperatures by considering and modifying negative brightness maps, which is similar in spirit to demanding positive phase curves and brightness maps when fitting the data. The mapping approach yielded a nightside temperature trend consistent with that of the disk-integrated approach.

\begin{figure*}
\centering
\resizebox{14.0cm}{!}{\includegraphics{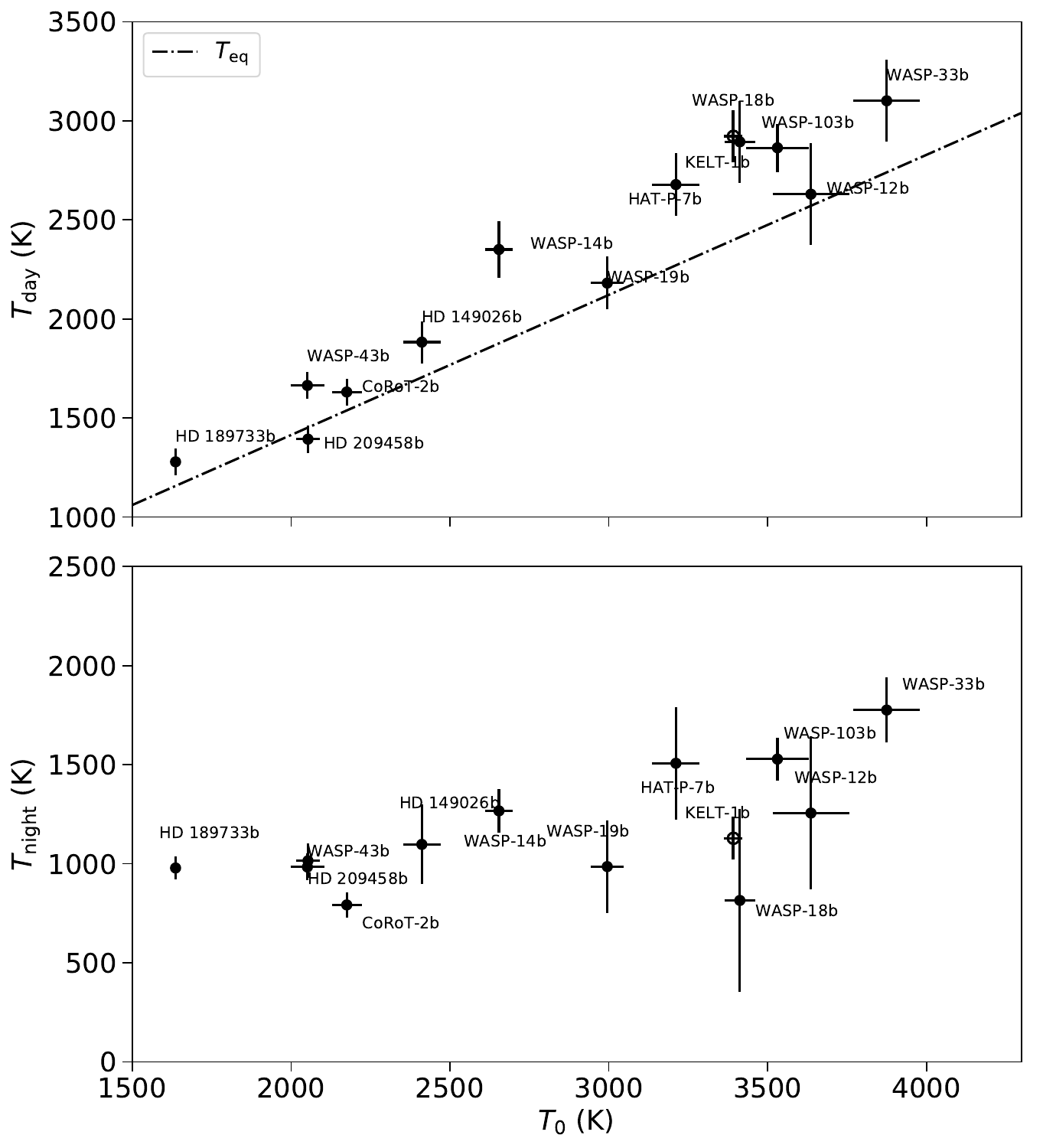}}
\caption{Dayside and nightside effective temperatures for twelve hot Jupiters, and one brown dwarf (KELT-1b). \textit{Top panel:} Dayside temperatures for the hot Jupiters in our analysis are proportional to the planets' irradiation temperatures, $T_{0} \equiv T_{\star}\sqrt{R_{\star}/ a}$, where $T_{\star}$ is the stellar effective temperature, $R_{\star}$ is the stellar radius, and $a$ is semi-major axis. The error bars correspond to the 1$\sigma$ confidence intervals. To guide the eye, we plot the equilibrium temperature $T_{\rm eq} \equiv (1/4)^{1/4} T_{0}$. \textit{Bottom panel:} Nightside effective temperatures. The error bars correspond to the 1$\sigma$ confidence intervals. Nightside temperatures are all around 1100K with a slight upward trend. \label{fig:nightside}}
\end{figure*}

In \Cref{fig:nightside} we show the dayside and nightside effective temperatures plotted against the stellar irradiation temperature, $T_{0} \equiv T_{\star}\sqrt{R_{\star}/ a}$, were $T_{\star}$ is the stellar effective temperature, $R_{\star}$ is the stellar radius, and $a$ is semi-major axis. The nightside temperatures are all around 1100K and exhibit a slight upward trend with stellar irradiation.  We tabulate the dayside temperature, nightside temperature, Bond Albedo, and heat recirculation efficiency for each planet in \Cref{table:results}. While this paper was under review, a similarly flat trend for nightside brightness temperature was reported \cite{Beatty2018}.  

Various theories have suggested that reradiation \cite{Showman2002}, advection, wave propagation \cite{Komacek2016}, molecular dissocation \cite{Bell2018}, coriolis forces, and magnetic drag could all play a role in atmospheric circulation on hot Jupiters. Models predict that the amount of day-night heat recirculation depends sensitively on planetary properties and the amount of stellar irradiation each planet receives, which vary between individual hot Jupiters \cite{Komacek2017,Komacek2018,ZhangShowman2017}.  

We fit the nightside temperatures using two models of atmospheric heat transport. The qualitative behaviour of each model is shown in Supplementary Figures 5 and 6. The first model is a semi-analytic energy balance model incorporating atmospheric radiation and advection \cite{Cowan2011a}, and predicts nightside temperatures given the planetary and stellar properties. We fit for two parameters: a common wind velocity, and P/g, where the latter quantity is the mass per unit area of the active layer of the atmosphere, that is, the layer that responds to instellation. The model was updated recently to include the effects of hydrogen dissociation and recombination, by solving the Saha equation to determine the amount of hydrogen dissociated at a given atmospheric temperature and the resulting heat sinks/sources \cite{Bell2018}. Hydrogen dissociation and recombination is predicted to significantly increase heat transport in ultra-hot Jupiters \cite{Bell2018,Komacek2018}.  

The second model is an analytic, dynamical model incorporating radiation, advection, magnetic drag, coriolis forces, and gravity waves \cite{Komacek2016}. This model predicts---rather than prescribes---the wind velocities for each hot Jupiter, and was shown to qualitatively match predictions of day-night temperature contrast from general circulation models. The model was recently updated to include the effects of hydrogen dissociation, albeit in a greatly simplified form \cite{Komacek2018}. As the magnetic field strengths of hot Jupiters are unknown to orders of magnitude \cite{Yadav2017}, we chose to neglect magnetic drag, but note that magnetic drag can potentially depress day-night heat transport for planets with very strong magnetic fields ($\sim 100$ G). We fit for a universal P/g, hence this model has only one fit parameter.  

\begin{figure*}
\centering
\resizebox{14.0cm}{!}{\includegraphics{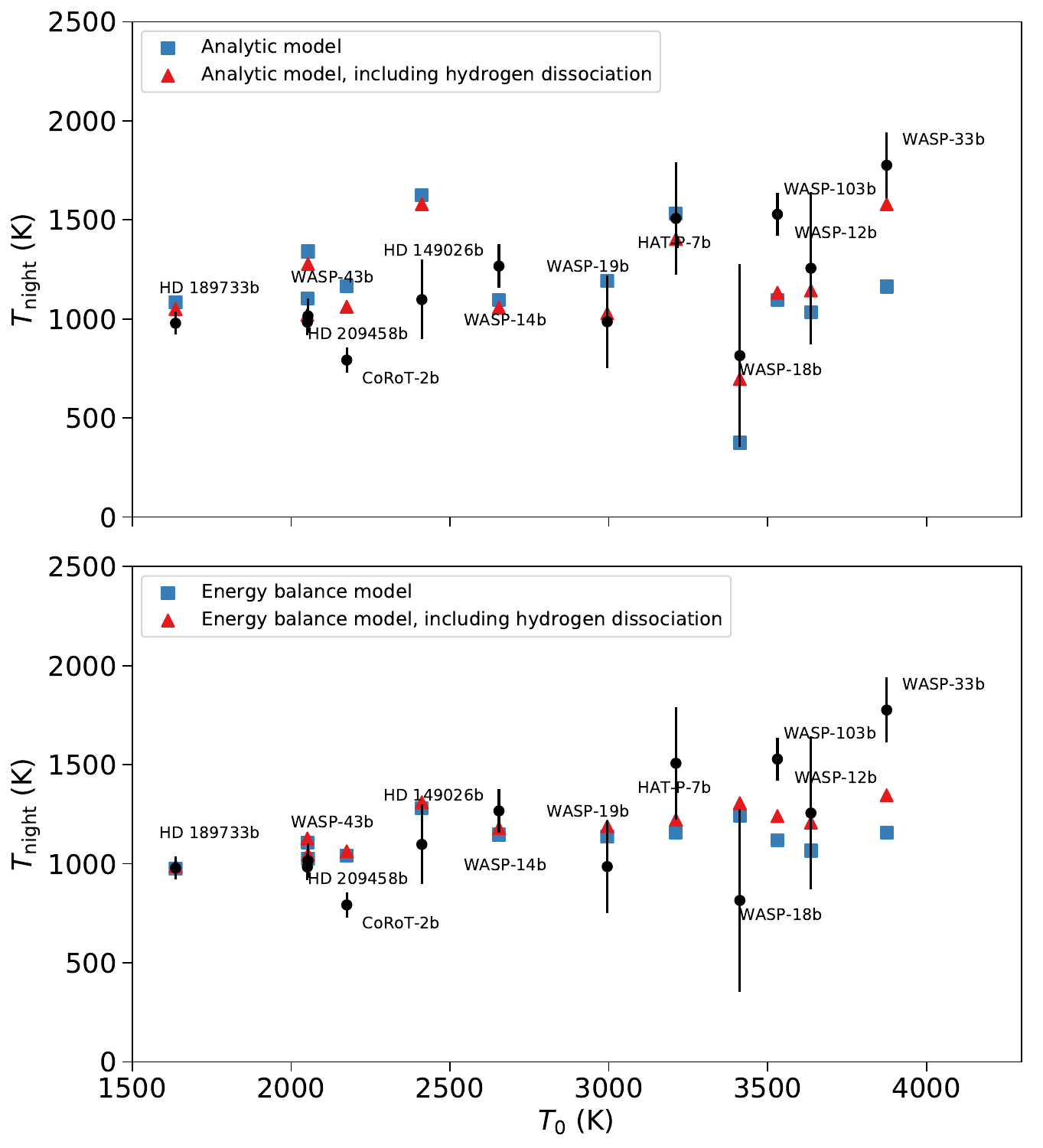}}
\caption{Best-fit models for the nightside temperatures of twelve hot Jupiters.  \textit{Top panel:} Best fit analytic, dynamical models \cite{Komacek2016,Komacek2018,ZhangShowman2017}. The error bars correspond to the 1$\sigma$ confidence intervals. The model including hydrogen dissociation is a much better fit ($\chi ^{2}$/datum $= 4.5$) than the model without ($\chi ^{2}$/datum $= 8.5$), but in both cases the model predicts greater planet-to-planet variance due to differences in predicted wind speeds. The wind speed depends on the gravity wave propagation timescale, which itself depends on the radius and mass of a planet. Differences in radius, mass, and rotation rate of these planets lead to variance in the predicted nightside temperatures.   \textit{Bottom panel:} Best fit semi-analytic energy balance models, using a common wind speed for all twelve planets \cite{Cowan2011a,Bell2018}. The error bars correspond to the 1$\sigma$ confidence intervals. The model that includes hydrogen dissociation is a better fit ($\chi ^{2}$/datum $= 3.4$) than the one without ($\chi ^{2}$/datum $= 4.3$), with $\Delta \chi^{2}=11$. The energy balance model, with a common wind speed for all twelve planets, is a better fit to the data than the analytic model, but neither fully captures the trend.  \label{fig:fitsCombined}}
\end{figure*}

For each model fit, we performed a grid search in parameter space to find the parameters that minimize $\chi^{2}$.  Models that allow for hydrogen dissociation provide better fits to the data than those without, even though this does not increase the number of parameters. The best fit model predictions can be seen in \Cref{fig:fitsCombined}. The semi-analytic energy balance model incorporating hydrogen dissociation yielded the best fit of all the models we considered. In the context of the energy balance model, the trend in observed nightside temperatures suggests that all hot Jupiters have similar wind velocities, contrary to predictions.

Alternatively, dynamical predictions may be correct, but ultimately overshadowed by optically thick nightside clouds. Clouds are predicted to be present on the nightsides of all hot Jupiters \cite{Parmentier2016,Powell2018,Roman2018}, but the cloud composition depends on the temperature, pressure, and cloud formation physics. Observationally, nightside clouds have been previously invoked to explain non-detections of nightside flux \cite{Kataria2015,Wong2015,Stevenson2017}. The nightside temperature trend \--- or lack thereof \--- implies that the hot Jupiters in our study all have nightside clouds that emit at similar temperatures. Vertical mixing sets the cloud top pressure, so in principle we could be seeing cloud tops from different cloud species that all happen to have similar vertical cloud-top temperatures.

\begin{figure*}
\centering
\resizebox{12.0cm}{!}{\includegraphics{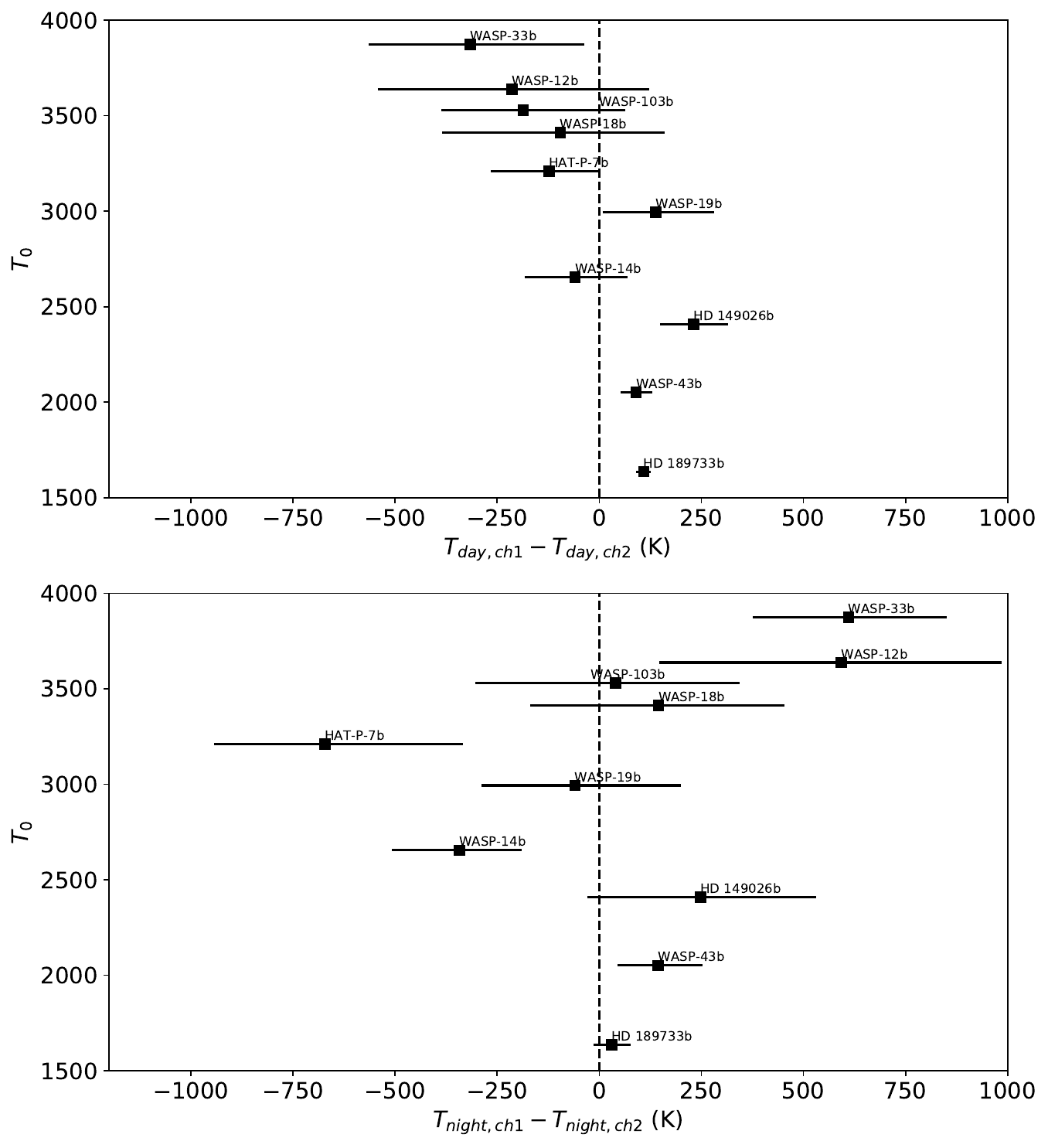}}
\caption{Difference in brightness temperatures at \textit{Spitzer} wavelengths 3.6~$ \mu$m and 4.5~$\mu$m for the ten planets with both 3.6~$ \mu$m and 4.5~$\mu$m phase curves. The error bars correspond to the 1$\sigma$ confidence intervals. \textit{Top Panel:}  Dayside brightness temperature colours. Cooler planets have blue IRAC colours (consistent with H$_{2}$0 absorption), while hotter planets are isothermal or slightly red. \textit{Bottom Panel:} Nightside brightness temperature colours. Half of the planets have nightside brightness temperature colours consistent with zero, meaning their nightside brightness temperatures are similar at both wavelengths. The rest are within $2\sigma$ of zero.
\label{fig:DeltaT}}
\end{figure*}

A simpler explanation is that hot Jupiters all have the same species of nightside clouds, which condense at a similar cloud-base temperature. The emitting temperature corresponds to the cloud-top temperature, which would be slightly cooler than the condensation temperature. These clouds would emit thermal radiation around the same effective temperature, and block outgoing longwave radiation from below, requiring clouds with large grains. Potential cloud species include manganese sulfide or silicate clouds, based on condensation curves \cite{Parmentier2016}.  As we show in \Cref{fig:DeltaT}, the nightside infrared colours are roughly isothermal. The similarity of the brightness temperature between \textit{Spitzer} bandpasses implies that they are probing parts of the atmosphere with similar temperatures, consistent with optically thick clouds.

Incorporating radiative feedback and detailed cloud microphysics is computationally intensive, which is the reason many studies have used cloud-free general circulation models, and post-processed clouds afterwards using the resulting temperature-pressure profiles and cloud condensation curves. However, post-processing of exoplanet clouds can lead to different predictions of cloud coverage, phase offsets, and day-night temperature contrasts than more intricate models \cite{Roman2018,Powell2018}. Fully three-dimensional models incorporating realistic cloud physics and heat transport due to hydrogen chemistry are clearly needed in order to properly understand hot Jupiters spanning the full range of irradiation temperatures.  Realistic treatments of magnetic effects may be necessary for the hottest planets \cite{Arcangeli2019,Yadav2017}. On the observational front, spectroscopic phase curve observations at longer wavelengths \cite{Morley2017}, with the Mid-Infrared Instrument onboard the \textit{James Webb Space Telescope}\cite{Bean2018}, and with the \textit{Atmospheric Remote-sensing Infrared Exoplanet Large-survey}\cite{Tinetti2018}, will make it possible to characterize the dominant cloud species on hot Jupiter nightsides. 

\begin{table*}
\centering
\caption{Dayside temperatures, nightside temperatures, and energy budget parameters for twelve hot Jupiters. We also include the brown dwarf KELT-1b. Our heat recirculation parameter, $\varepsilon$, ranges from 0, for no day-night heat recirculation, to 1, for perfect day-night heat recirculation \cite{Cowan2011b}. $A_{\rm B}$ is the Bond Albedo. \label{table:results}}
\begin{tabular}{llllll}
 & $T_{0}$ K    & $T_{day}$ K  & $T_{night}$ K & $A_{B}$ &$\varepsilon$ \\
 \hline \\ [-1.5ex]
HD 189733b & 1636$\pm$14  & 1279$\pm$68  & 979$\pm$58    & 0.16$^{0.11}_{0.1}$  & 0.59$^{0.12}_{0.11}$ \\
WASP-43b   & 2051$\pm$53  & 1664$\pm$69  & 984$\pm$67    & 0.22$^{0.13}_{0.12}$ & 0.27$^{0.07}_{0.07}$ \\
HD 209458b & 2053$\pm$38  & 1393$\pm$70  & 1015$\pm$86   & 0.52$^{0.08}_{0.09}$ & 0.51$^{0.15}_{0.13}$ \\
CoRoT-2b   & 2175$\pm$47  & 1631$\pm$67  & 792$\pm$64    & 0.48$^{0.09}_{0.1}$  & 0.13$^{0.05}_{0.04}$ \\
HD 149026b & 2411$\pm$59  & 1883$\pm$106 & 1098$\pm$201  & 0.33$^{0.14}_{0.15}$ & 0.25$^{0.18}_{0.13}$ \\
WASP-14b   & 2654$\pm$43  & 2351$\pm$142 & 1267$\pm$111  & 0.12$^{0.14}_{0.08}$ & 0.22$^{0.09}_{0.06}$ \\
WASP-19b   & 2995$\pm$52  & 2181$\pm$133 & 986$\pm$233   & 0.54$^{0.1}_{0.12}$  & 0.1$^{0.12}_{0.07}$  \\
HAT-P-7b   & 3211$\pm$75  & 2678$\pm$158 & 1507$\pm$285  & 0.21$^{0.16}_{0.14}$ & 0.22$^{0.17}_{0.12}$ \\
KELT-1b    & 3391$\pm$29  & 2922$\pm$132 & 1128$\pm$108  & 0.18$^{0.12}_{0.11}$ & 0.06$^{0.03}_{0.02}$ \\
WASP-18b   & 3412$\pm$49  & 2894$\pm$206 & 815$\pm$463   & 0.26$^{0.17}_{0.15}$ & 0.01$^{0.07}_{0.01}$ \\
WASP-103b  & 3530$\pm$99  & 2864$\pm$122 & 1528$\pm$108  & 0.27$^{0.12}_{0.14}$ & 0.19$^{0.06}_{0.05}$ \\
WASP-12b   & 3636$\pm$121 & 2630$\pm$258 & 1256$\pm$386  & 0.53$^{0.16}_{0.2}$  & 0.13$^{0.23}_{0.1}$  \\
WASP-33b   & 3874$\pm$104 & 3101$\pm$206 & 1776$\pm$165  & 0.28$^{0.16}_{0.16}$ & 0.25$^{0.11}_{0.08}$\\
\hline \\ [-1.5ex]
\end{tabular}
\end{table*}

\newpage

\newpage

\section*{\large{Acknowledgements}}
The authors thank Taylor Bell for providing the updated energy balance model code, Tad Komacek for providing his analytic day-night temperature difference code, and to Laura Kreidberg for an early look at her WASP-103b phase curve paper. Thanks to Jo{\~ a}o Mendon{\c c}a for providing the phase curve parameters from his WASP-43b paper. Thanks to Emily Pass for an early look at her Gaussian process temperature estimate results. Thanks to Jacob Bean and Vivien Parmentier for helpful discussion and comments about the manuscript.  
\section*{\large{Author Information}}
\subsection{Contributions}
D.K. led the data analysis, and wrote the manuscript. N.B.C. discussed ideas and contributed to writing the manuscript. L.D. provided the \textit{Spitzer} data analysis pipeline and helped with reducing \textit{Spitzer} phase curves.

\subsection{Corresponding author}
Correspondence and requests for materials should be addressed to Dylan Keating (dylan.keating@mail.mcgill.ca).

\newpage
\section*{\large{Methods}}\vspace{-4mm}
We estimated nightside effective temperatures with two different methods. We outline the methods in the sections that follow.

\section*{Method 1: Disk-Integrated Flux}\vspace{-4mm}
Our fiducial analysis used the disk-integrated flux, from phase curves, to estimate effective temperatures. Previous efforts have used weighted averages or linear interpolation \cite{Cowan2011b, Schwartz2015,Schwartz2017}. We used Gaussian Process regression (GP), to estimate the bolometric flux, and subsequently  effective temperature and uncertainty given a handful of brightness temperatures, as it has recently been shown to produce more accurate uncertainty estimates (work in prep, Pass et. al 2019, submitted to MNRAS).  

The disk-integrated nightside brightness temperature is given by 
\begin{equation}
T _ { \rm b,night } ( \lambda ) = \frac { h c } { \lambda k } \left[ \ln \left( 1 + \frac { e ^ { h c / \lambda k T _ { * } } - 1 } { F _ { \rm night} / \delta _ { \rm tra } } \right) \right] ^ { - 1 },
\end{equation}
where $\lambda$ is the wavelength of the observation, $T_{\star}$ is the brightness temperature of the star at that wavelength, $\delta _ { \rm tra }$ is the transit depth, and $F_{\rm night}$ is the planet-to-star flux ratio at a phase angle of $\pi$, where phase angle is defined to be $0$ at secondary eclipse.

\subsection{Method 1: Propagation of Uncertainties}
As the most common \textit{Spitzer} decorrelation techniques have been shown to produce accurate, reproducible results \cite{Ingalls2016}, we chose to take all positive phase curves at face value. To estimate uncertainties on each planet's disk-integrated nightside flux and brightness temperature, we propagated uncertainties on planetary and stellar properties, and phase curve parameters, using a 1000 step Monte Carlo. The relevant physical properties are: the stellar effective temperature, stellar surface gravity, stellar metallicity, transit depth, and ratio of semi-major axis to stellar radius. We took the most up-to-date values from the literature. For each draw, we randomly sampled each parameter from a Gaussian centered on each best-fit published value, with the width given by the published  uncertainty. This gives an approximately Gaussian probability density function for the nightside flux ${\rm Prob}(F_n) = f(T_n),$
where $F_n$ is the nightside flux.

To calculate the brightness temperature for each nightside flux value, we inverted the Planck function at each flux to obtain a probability density function for nightside brightness temperature, $T_{n}$. This can be thought of as transforming the nightside flux probability density function to a function of nightside temperature through a change of variables. We have,
\begin{equation}
    {\rm Prob}(T_n) = g(F_n(T_n)) \frac{dF_n}{dT_n},
\end{equation}
where $F_n(T_n)$ is the Planck function (as a function of temperature, holding wavelength fixed), and $\frac{dF_n}{dT_n}$ is the derivative of the Planck function with respect to temperature. This transformation is only defined for positive fluxes and temperatures.

For most of the planets in our study, the nightside flux distribution is well above zero. The nightside temperature probability distribution also has a Gaussian-like shape, so we used the peak and width for our best-fit and uncertainty values. We took the average of the upper and lower limits when using the brightness temperatures to infer effective temperatures.  

 For planets with low or negative nightside flux, parts of the nightside flux probability distribution do not correspond to physical temperatures. This is typically interpreted to be a strong non-detection of nightside flux. Mathematically this is allowed, but physically, negative fluxes and temperatures are impossible. An example is HAT-P-7b at $3.6\mu$m, where the peak of the probability density function ${\rm Prob}(F_n)$ is negative. In this case we set the best-fit flux, and hence brightness temperature, to zero, and used the width of the nightside flux distribution to calculate a $1 \sigma$ upper limit on the brightness temperature, which we used as the error when estimating the bolometric flux. For planets with small but non-zero nightside flux (like WASP-18b at $3.6\mu$m), a significant part of the flux distribution is negative, and the lower part gets truncated when converting to errors in brightness temperature. In these cases, we used the upper limit on brightness temperature when estimating bolometric flux and effective temperature, which is more conservative than taking the average of the upper and lower limits.

\subsection{Method 1: Brightness Temperature Difference Plot}
To generate \Cref{fig:DeltaT}, we used a 1000 step Monte Carlo. For each step in the Monte Carlo we calculated the difference between the 3.6~$\mu$m and 4.5~$\mu$m brightness temperatures. We took the mean and standard deviation of the distribution of differences for each planet.

\section*{Method 2: Mapping Method}\vspace{-4mm}
We also calculated dayside and nightside temperatures by considering the brightness maps implied by each phase curve. For a planet on a circular, edge-on orbit, its orbital phase curve can be analytically inverted into a longitudinal brightness map \cite{Cowan2008}. WASP-14b has the highest eccentricity of the sample, $e=0.08$. General circulation models using a small eccentricity ($e=0.15$) predict negligible differences in circulation patterns compared to circular orbits \cite{Lewis2010}. For our purposes we treated the orbits of WASP-14b and the lower eccentricity planets in our sample as circular. We defined the phase curves to be $F(\xi),$ where $\xi$ is the planet's phase angle ($\xi = 0$ at secondary eclipse, $\xi = \pi$ at transit). The corresponding brightness maps are defined as $J(\phi)$,  where $\phi$ is longitude from the substellar point. We set $F(\xi=0)$ equal to the eclipse depths, and obtained the map parameters analytically \cite{Cowan2008}.

Phase curves provide weak constraints on North-South asymmetry of planets \cite{Cowan2013,Cowan2017}. It is possible to determine the latitudinal distribution using eclipse mapping, but so far this has only been done for HD 189733b at 8$~\mu$m \cite{Majeau2012,DeWit2012,Rauscher2018}. We therefore marginalize over the uncertainty in latitudinal brightness distributions when constructing the two-dimensional bolometric flux maps. 

From the bolometric flux maps for the twelve planets, we obtained an estimate of the dayside and nightside effective temperatures of each planet. 

\subsection{Method 2: Latitudinal Brightness Profiles}
Longitudinal maps, $J(\phi)$, are weighted by the visibility of the observer, since the phase curve measures the disk-integrated flux from the planet. For an equatorial observer (a zero-obliquity planet orbiting edge-on), the longitudinal maps are related to the two-dimensional brightness distribution as a function of planetary co-latitude and longitude,
$I(\theta,\phi)$, by 
\begin{equation}\label{eq:1}
J(\phi) = \int_{0}^{\pi}I(\phi,\theta) \sin^{2}\theta d \theta. 
\end{equation}
One of the powers of sine comes from the area element in spherical coordinates, and the other comes from the visibility for an equatorial observer. The longitudinal map $J(\phi)$ effectively integrates over the latitudinal dependence of $I(\phi,\theta)$. We adopted the simplifying assumption that $I(\phi,\theta)$ is separable, and accounted for our ignorance of the latitudinal dependence of brightness by letting it vary as $\sin ^{\gamma}\theta$ with a polar brightness $I_{\rm pole}$. The expression is 
\begin{equation}\label{eq:2}
I_{\lambda}(\phi,\theta)= \left(\frac{J_{\lambda}(\phi) - \pi I_{\rm pole}/2}{\int_{0}^{\pi} \sin^{2+\gamma} \theta d \theta}\right)\sin^{\gamma}\theta+I_{\rm pole}(1-\sin^{\gamma}\theta),
\end{equation} where $I_{\rm pole}$ is a constant representing the intensity at the poles. The full derivation can be found at the end of the Methods section. The $\gamma = 0$ case represents perfect poleward heat transport, or a constant temperature in the latitudinal direction. In the Rayleigh-Jeans limit of long wavelength, $I(\theta)\propto T(\theta)$, and thus $I(\theta) \propto T(\theta) \propto \sin^{1/4}(\theta)$ for no poleward heat transport, that is, $\gamma=1/4$. To be conservative, we drew samples from the range $0 < \gamma \leq 1$, as we find that the value of $\gamma$ doesn't drastically affect our calculated quantities. In Supplementary Figure 8 we show how the value of $\gamma$ changes the latitudinal brightness profile.  

The brightness temperature map is related to the intensity map by the inverse Planck function
\begin{equation}\label{eq:3}
T_{\lambda}(\phi,\theta) =\frac{hc}{\lambda k}\left[ {\ln} \left(1+\frac{(e^{hc/\lambda k T_{\rm *}}-1)(R_{p}/R_{\star})^{2}}{\pi I_{\lambda}(\phi,\theta)}\right)\right]^{-1},
\end{equation}
where $T_{\rm *}$ is the brightness temperature from Phoenix stellar models \cite{Phoenix}.

\subsection{Method 2: Brightness Temperatures to Effective Temperatures}
From the wavelength dependent brightness maps in \Cref{eq:3}, we inferred effective temperature maps. If the full spectrum at each location was known, one could integrate it to get the effective temperature at each location. Instead, we must estimate the bolometric flux by interpolating between, and extrapolating from, a few brightness temperatures. The Gaussian process regression used for the disk-integrated analysis is too computationally expensive to use at each location on the planet. We instead approximated the effective temperature via the error weighted mean \cite{Schwartz2015,Schwartz2017} of the brightness temperatures (or, in practice, the arithmetic mean of brightness temperatures, but embedded in a Monte Carlo). We adopted systematic uncertainties calibrated by performing such estimates on synthetic spectra (work in prep, Pass et. al 2019). We took the arithmetic mean of the individual brightness temperatures at each location as an estimate of the effective temperature,
\begin{equation}
T_{\rm eff}(\phi,\theta) = \frac{1}{n} \sum_{n}T_{\lambda,n}(\phi,\theta),
\end{equation} where $n$ is the number of wavelengths. We propagated errors in a Monte Carlo fashion. From $T_{\rm eff}(\phi,\theta)$ we calculated the disk-integrated dayside and nightside effective temperatures using the Stefan-Boltzmann law,
\begin{equation}
T_{\rm day} = \frac{1}{2\pi} \int_{-\pi/2}^{\pi /2}\int_{0}^{\pi}T_{\rm eff}^{4}(\phi,\theta) \sin \theta d \theta d \phi,
\end{equation}
and
\begin{equation}
T_{\rm night} = \frac{1}{2\pi} \int_{\pi/2}^{-\pi /2}\int_{0}^{\pi}T_{\rm eff}^{4}(\phi,\theta) \sin \theta d \theta d \phi.
\end{equation}

We also calculated the Bond albedo, the fraction of incoming stellar power that the planet reflects to space,
\begin{equation}\label{eq:5}
A_{\rm B}\equiv 1-\frac{\oint T_{\rm eff}^{4}(\phi,\theta) \sin \theta d \theta d \phi}{\pi T_0^{4}}.
\end{equation}

For day-night heat recirculation, we computed the ratio of heat radiated by the nightside to the total heat radiated by the planet,
\begin{equation}\label{eq:6}
\mathcal{P}  = \frac{\int_{\pi/2}^{-\pi /2} \int_{0}^{\pi} T_{\rm eff}^{4}(\phi,\theta) \sin \theta d \theta d \phi}{\oint T_{\rm eff}^{4}(\phi,\theta) \sin \theta d \theta d \phi}.
\end{equation} Since the nightside absorbs no stellar radiation, this is the amount of heat that has moved from the dayside to nightside. A value of zero implies that no heat is transported to the nightside, and a value of 0.5 implies that half of the absorbed incoming stellar flux is recirculated to the nightside. 

For each planet, we calculated $T_{\rm day}$, $T_{\rm night}$, $A_{\rm B}$, and $P$ simultaneously for $10^{5}$ steps of a Monte Carlo. At each step we randomly drew all measured physical planetary parameters and published phase curve parameters from Gaussian distributions centered around their published values, with standard deviation given by their published uncertainties. We marginalized over our uncertainty of $\gamma$ and $I_{\rm pole}$ by drawing these from a random uniform distribution where $0 \leq \gamma \leq 1$ and $0 \leq I_{\rm pole} \leq  \min \frac{2 J(\phi)}{\pi}$. The constraint on $I_{\rm pole}$  ensures that the poles are not hotter than the equator.  Lastly, we added the systematic uncertainty associated with estimating effective temperatures using the mean of a limited number of brightness temperatures; the systematic uncertainties are esimated based on retrieval exercises with synthetic cloud-free dayside emission spectra. The 1$\sigma$ systematic uncertainties are $23\%$ for planets with a phase curve at just 4.5~$\mu$m, $13\%$ for planets with phase curves at 3.6~$\mu$m and 4.5~$\mu$m, and $3\%$ for planets with phase curves at 3.6~$\mu$m, 4.5~$\mu$m, and 1.4~$\mu$m. This is a conservative estimate, as the observed nightside brightness temperatures are closer to isothermal than the dayside brightness temperatures (\Cref{fig:DeltaT}).

\subsection{Method 2: Sensitivity Analysis}

To determine how much each measured parameter affects the overall error of the calculated values, we performed a sensitivity analysis using the measured values, varying one parameter at a time. For most planets, the biggest source of uncertainty in the nightside temperatures is the systematic error we introduced, followed by the $I_{\rm pole}$ term. This suggests that obtaining phase curves at more wavelengths, as well as eclipse mapping, will yield better estimates of nightside effective temperatures. 

\subsection{Phase Curves and Brightness Maps} \label{sec:negbright}
The first exoplanet map was of HD 189733b at 8$~\mu$m \cite{Knutson2007}. It showed an eastward shifted hotspot on the planet, in line with theoretical predictions of equatorial, super-rotating jets \cite{Showman2002}. With the exception of HD 189733b \cite{Knutson2007,Knutson2009}, 55 Cancri e \cite{Demory2016,Angelo2017}, CoRoT-2b \cite{Dang2018}, and WASP-43b \cite{Spiderman}, WASP-103b \cite{Kreidberg2018}, and KELT-1b \cite{Beatty2018}, most phase curves have been fit and published without considering the brightness maps that could have produced them. 

We distinguish between two problematic cases: negative phase curves, and positive phase curves that imply negative brightness maps, and explain how we handle these problematic cases. We summarize the suite of phase curves in Supplementary Table 3. 

\subsection{Negative Phase Curves}
A phase curve that is negative at any value of orbital phase guarantees that the underlying brightness distribution (map) is negative at some longitudes, because a phase curve measures the disk-integrated flux. Every phase curve that goes negative at any point implies the planet has negative flux somewhere.

The planets HAT-P-7b, WASP-14b, and WASP-43b have published phase curves that are negative on their nightsides, which ensures unphysical, negative brightness maps for these planets. The best one can do without refitting the phase curves is to modify the phase curves or brightness maps in some way. A possible solution previously adopted for WASP-43b is simply setting negative regions of each of brightness maps to zero \cite{Keating2017}. The WFC3 phase curves for WASP-43b have since been refit while enforcing physically possible brightness maps and accounting for reflected light, resulting in a much higher nightside temperature than previous reported \cite{Stevenson2014,Spiderman}. The \textit{Spitzer} phase curves for WASP-43b were refit by using a different instrument sensitivity model, shown to be better at removing residual red noise due to intra-pixel sensitivity, also resulting in much higher nightside temperatures \cite{Mendonca2018}. We use the reanalyzed \textit{Spitzer} phase curves for our analysis. For the WFC3 phase curve, we treated the negative nightside flux as an upper limit, rather than simply setting the negative parts of the map to zero. We do not use the reanalyzed WFC3 phase curves as they were not fit with sinusoids, and hence could not be treated in a consistent manner to the other phase curves.

The published HAT-P-7b and WASP-14b  $3.6~\mu$m phase curves are negative on their nightsides. This can occur when not enforcing positive brightness maps when fitting the data. We refit both phase curves while enforcing physically possible phase variations, using a polynomial function to model detector systematics \cite{Dang2018}. For WASP-14b, we were able to obtain a good fit. The nightside temperature we infer is 4 K lower than when using the Monte Carlo rejection method. For HAT-P-7b, we were not able to obtain a good fit without allowing the phase curve to have significantly negative nightside flux. As the planet cannot have negative brightness, this could potentially be due to some unmodelled stellar effect, such as non-uniform stellar brightness, or that the detector models are inadequate for these particular data. For the purposes of this study, we chose to treat the negative nightside flux as an upper limit when estimating the effective temperature.     

\subsection{Positive Phase Curves, Negative Brightness Maps}
It is also possible to measure a strictly positive phase curve, yet infer a brightness map that is not strictly positive. This is the case for WASP-12b, WASP-18b, WASP-19b, and WASP-103b. Although it is \emph{mathematically} possible to obtain a  non-negative phase curve from a brightness map that is not strictly positive, such a brightness map is physically impossible. This was pointed out long ago in the case of reflected light curves from asteroids \cite{Russell1906}, which is mathematically similar to the thermal emission case. Brightness maps obtained from inverting sinusoidal phase curves are not unique, as there is a nullspace of the transformation from map to light curve--- excluding the fundamental mode, any odd sinusoidal mode present in the brightness map of a synchronously rotating planet on a circular, edge-on orbit will integrate to zero over a hemisphere, and will thus be invisible in the phase curves \cite{Cowan2008,Cowan2013}. If a measured phase curve implies a negative brightness map, then it may be possible to add higher order odd harmonics to correct the map --- indeed, if a solution exists, then odd brightness map harmonics are necessary to ensure a physically possible solution.

For example, WASP-18b has strictly positive phase curves that were fit with first and second order sinusoids \cite{Maxted2013}. However, the published $3.6~\mu$m phase curve parameters imply a negative brightness map at this wavelength. For each draw in our Monte Carlo, if the phase curve is positive but the brightness map is negative at any location for any planet, we numerically solve for the smallest amplitude third order harmonic that makes the brightness map non-negative. If no such solution exists, we reject the draw. We demonstrate this in Supplementary Figure 4.

The brown dwarf KELT-1b also has positive phase curves that imply negative brightness maps \cite{Beatty2018}. The authors showed that a smoothed trapezoidal brightness map integrates to give an approximately sinusoidal phase curve close to their fiducial phase curve for KELT-1b, and conclude that KELT-1b's map must be non-sinusoidal. They argue that this solves the problem of negative brightness maps, and implies that all planets with seemingly negative sinusoidal brightness maps must instead have non-sinusoidal maps. As we have shown, for some planets with positive phase curves, the brightness map can be made non-negative by just adding the third harmonic to the brightness map. In fact, adding higher order sinusoids allows for trapezoidal temperature maps, or any other continuous function (in other words, Fourier analysis).  The odd harmonic method is elegant and more robust than adopting a specific non-sinusoidal parameterization, and does not alter the phase curve. It may be necessary to fit for the odd map harmonics when fitting phase curves\--- even though they are not visible in the phase curve, they may be needed to ensure a physically possible map.

Lastly, the phase curves of WASP-12b are contentious --- if the fiducial, polynomial fit for the 4.5$~\mu$m phase curve is taken to be solely due to brightness variations of a spherical planet, the map is negative and unphysical \cite{Cowan2012}. The authors note that part of the second harmonic could be due to ellipsoidal variations. A reanalysis of the same data, and a second set of phase observation at the same wavelengths, found that the 4.5$~\mu$m results were consistent with the previous results \cite{Zhang2018}. We adopt the interpretation ultimately chosen by the authors of the first paper: some of the second harmonic in the 4.5$~\mu$m phase curve is due to the planet's inhomogeneous temperature map, but the rest is due to ellipsoidal variations. To be consistent with their interpretation, we set the planet's aspect ratio to 1.5, calculated the resulting amplitude \cite{Cowan2012}, and subtracted it from the second order amplitude to yield a non-negative brightness map.

\subsection{Dynamical Model}
The radiative timescale in the analytic, dynamical model is scaled by pressure at the base of the radiatively active layer of the atmosphere, and equilibrium temperature \cite{Komacek2017,Komacek2018,ZhangShowman2017}. Supplementary Figure 5 shows a version of the model where all the planets have the same physical properties, but the irradiation temperature varies. 

We updated the radiative timescale formulation to scale with $P/g$, as with the energy balance model. We neglected magnetic drag.  Magnetic drag could decrease nightside temperatures for planets where magnetic drag is predicted to be significant ($T_{0} \sim 2000$~K and up) \cite{Komacek2017}. Presumably, hotter planets have more ionized atmospheres, and thus shorter drag timescales due to interactions with magnetic fields. We ``anchor'' the air column mass to the nightside temperature of HD 189733b which has $T_{0}=1636$~K, and thus presumably no appreciable magnetic drag, so we can safely set the magnetic drag timescale to infinity. The predictions for the nightside temperatures of the more irradiated planets at this air column mass are all significantly \textit{lower}, rather than higher, than the observed nightside temperatures. See Supplementary Figure 7. Nightside clouds could only further depress the nightside effective temperature. This simple parameterization of magnetic drag is not sufficient to explain the entire nightside temperature trend, and including it gives a worse fit to the observations. This does not allow us to exclude the effects of magnetic fields in hot Jupiter atmospheres--- instead, it motivates the need to include magnetohydrodynamics in general circulation models of hot Jupiter atmospheres. 

The dynamical models do not predict the dayside or nightside temperatures themselves, but rather the day-to-night temperature contrast. To predict the nightside temperature, we use the analytic expression from the model to calculate the day-night temperature contrast, and solve for the nightside temperature, assuming that the dayside temperature is equal to the equilibrium temperature defined by $T_{\rm eq} = (1/4)^{1/4} T_{0}$ (a good approximation, as shown in the top panel of \Cref{fig:nightside}). In true radiative equilibrium, the nightside temperature of a tidally locked, synchronously rotating planet would be zero. However, we note that GCMs suggest that the nightsides of hot Jupiters can never reach temperatures as cold as expected in radiative equilibrium. \cite{Komacek2017}.

Rather than use a common photosphere pressure among the planets as has been previously done, we fit for a common air column mass above the emitting region, that is, the photosphere pressure ($P$) scaled by acceleration due to gravity ($g$): $P/g$. It is more a realistic assumption than a common photosphere pressure among the planets, as hot Jupiter masses, and hence surface gravities, can span an order of magnitude.

\section*{\large{Data availability}}
The data that support the plots within this paper and other findings of this study are available from the corresponding author upon reasonable request. 
\section*{\large{Code availability}}
The Gaussian process regression code used is publicly available, and can be found \href{https://github.com/ekpass/gp-teff}{here}. The \textit{Spitzer} Phase Curve Analysis pipeline is publicly available and can be found \href{https://github.com/lisadang27/SPCA}{here}.

\end{document}